%% file: discPrecision.tex
\documentclass[usenatbib,twocolumns]{mnras}
\usepackage{xcolor}
\pagecolor{white}
\usepackage{graphicx}
\usepackage{amsmath}
\usepackage{amssymb}
\usepackage{bm}
\usepackage{color}
\usepackage{physics}
\usepackage{subfigure}
\usepackage{accents}
\usepackage{yfonts}
\usepackage{mathrsfs}
\usepackage{textgreek}
\usepackage{comment}
\usepackage{makecell}
\usepackage{graphicx,color}
\usepackage{footnote}
\makesavenoteenv{tabular}
\input{macro}

\usepackage[normalem]{ulem}
\title[Precision with inhomogeneous stochasticity]
{
Influence of inhomogeneous stochasticity on the falsifiability of mean-field theories and examples from accretion disc modeling}

\author[Zhou \& Blackman]
{Hongzhe Zhou$^{1,2,3}$%
\thanks{Email address for correspondence: hongzhe.zhou@su.se},
Eric G. Blackman$^{2,3}$%
\thanks{Email address for correspondence: blackman@pas.rochester.edu}\\
$^1$
Nordita, KTH Royal Institute of Technology and Stockholm University,
Hannes Alfv\'ens v\"ag 12, SE-106 91 Stockholm, Sweden\\
$^2$ Department of Physics and Astronomy, University of Rochester, Rochester, NY, 14627, USA\\
$^3$ Laboratory for Laser Energetics, University of Rochester, Rochester NY, 14623, USA\\
}

\begin{document}

\date{}
\pagerange{\pageref{firstpage}--\pageref{lastpage}} \pubyear{}
\maketitle
\label{firstpage}

\begin{abstract}
Despite spatial and temporal fluctuations in turbulent astrophysical systems, mean-field theories can be used to describe their secular evolution.
However, observations taken over time scales much shorter than dynamical time scales capture a system in a single state of its turbulence ensemble.
Comparing with mean-field theory can falsify the latter only if the theory is additionally supplied with a quantified precision.
The central limit theorem provides appropriate estimates to the precision only when fluctuations contribute linearly to an observable and with constant coherent scales.
Here we introduce an error propagation formula that relaxes both limitations, allowing for nonlinear functional forms of observables and inhomogeneous coherent scales and amplitudes of fluctuations.
The method is exemplified in the context of accretion disc theories, where inhomogeneous fluctuations in the surface temperature are propagated to the disc emission spectrum--the latter being a nonlinear and non-local function of the former.
The derived precision depends non-monotonically on emission frequency.
Using the same method, we investigate how binned spectral fluctuations in telescope data change with the spectral resolving power.
We discuss the broader implications for falsifiability of a mean-field theory.
\end{abstract}

\section{Introduction}
Mean-field approaches have been widely applied in theoretical astrophysics including for models of stellar structure, convection \citep{CoxGiuli1968}, accretion discs \citep{SS73}, magnetic dynamos \citep{RobertsSoward1975,KrauseRaedler1980}, and interpreting large-scale structure in the Universe. 
In such approaches, physical quantities are typically decomposed into mean and fluctuating parts, and the former is explicitly solved for while the latter is modeled using closure methods.
To facilitate an analytically tractable theory, an infinite scale separation between the mean and fluctuating fields is often assumed, allowing for Reynolds averaging rules: linearity of averaging, interchangeability between averaging and differentiation, and invariance of mean quantities under subsequent averaging \cite[see, e.g.,][]{ZhouBlackmanChamandy2018}.

In contrast to this idealized assumption however, realistic astrophysical turbulent flows often exhibit finite and moderate scale separations. The ratio between the system and the turbulence scales is $\sim\mathcal{O}(10)$ to $\sim\mathcal{O}(100)$
 for geometrically thin accretion discs,
$\sim\mathcal{O}(10)$ for galaxies, $\sim\mathcal{O}(5)$ in the solar convection zone, and even lower for geometrically thick accretion discs.
Consequently, the assumption of infinite scale separation leads to at least two classes of discrepancies between theoretical predictions and observations \citep{ZhouBlackmanChamandy2018}:
First, mean-field equations need to be modified with correction to account for the violation of Reynolds rules because double averages are not necessarily equal to single averages. 
It follows that the \textit{accuracy} of the theory is threatened if equations derived assuming infinite scale separation are applied to flows with finite scale separation.
Secondly, finite scale fluctuations in observational data do not necessarily average to zero.
This poses an incompatibility if comparing to theoretical predictions that assume fluctuations have zero means because disagreement may be misinterpreted as systematic inaccuracy rather than stochastic imprecision. 

Previous work \citep{ZhouBlackmanChamandy2018} 
offered a solution to the latter problem, by theoretically estimating the influence from stochastic fluctuations coming from all scales below the mean-field scales, in turn defining the \textit{precision} of a mean-field theory. The deviations between observations and the theory can then be determined to be within or without the precision error bars of the theory, and a better assessment of whether they are stochastic or systematic can be deduced.

The question of computing precision of mean-field predictions can be mathematically formulated as follows:
Let $\mathcal{A}(\bm\tau)$ be some field
with 
(i) mean $A$ to predicted by some theory, 
(ii) fluctuations $\delta \mathcal{A}$ that modeled statistically; 
and $\bm\tau$ is a set of formal variables representing spatial coordinates, time, or any other parameters that $\mathcal{A}$ may depend on.
If $K$ is an observable related to $\mathcal{A}$ via
\beq
K=\int_{\mathcal V} k(\mathcal{A}(\bm\tau))\text{d}\bm\tau
\eeq
where $\mathcal V$ is some appropriate interval and $k$ is some function of $\mathcal{A}$, what is the fluctuation in $K$ due to that of $\mathcal{A}$?

\cite{ZhouBlackmanChamandy2018} focused on the context where $\mathcal{A}$ is a galactic magnetic field, $K$ is the Faraday rotation, and the integration is carried out over a line of sight.In this case, $k(\mathcal{A})$ is the density-weighted magnetic field along the line of sight, and thus a linear function of the mean magnetic field. The coherence length and the variance of the fluctuations were taken to be independent of $\tau$, the position along the line of sight.
The integral of $k$ above is then replaced by a discrete summation and the central limit theorem readily applies.
In the ensemble of different realizations of turbulence, the variance of $K$ is then simply $N$ times smaller than that of $k$ where $N$ is the number of turbulent eddies along the line of sight.

In the present work, we generalize the method to include cases where $k(\mathcal{A})$ is a nonlinear function of $\mathcal A$, and both the variance and the coherent scales of the fluctuations depend on $\bm\tau$.
We apply the formalism to accretion disc theories. For a wide range of accretion models, angular momentum transport is modeled by solving for the dynamics of the mean velocity field with angular momentum transport mediated by an imposed mean stress. These models are manifestly mean-field theories, and predictions from these models should therefore be presented with precision error bars.

Our method herein enables efficiently quantifying fluctuations about the mean-field predictions for the disc spectrum at all photon frequencies, thereby providing a quantitative way of accessing the precision of the models considered.
This is accomplished by propagating local disc perturbations with given amplitudes and coherent scales to the emission spectrum.
At a given photon frequency, this requires including contributions from the entire disc, superseding the local treatment of \cite{Blackman2010}.
Quantifying the theoretical precision 
 becomes particularly important when comparing to snapshot observations of a source because then the observed fluctuations cannot be identified by their temporal characteristics.

In Section \ref{sec:error_prop} we introduce the needed method for computing fluctuations of an integral from local contributions with inhomogeneous amplitudes and coherent scales.
In Section \ref{sec:spectrum} we apply the method to a generic thermal disc spectrum, 
and more specifically to 
protoplanetary discs in
 Section \ref{sec:pd}, and dwarf novae in \ref{sec:dn}. 
In Section \ref{sec:tel} we consider how the derived imprecision should be binned when taking into account comparison to observations with finite telescope resolving power.
We summarize in Section \ref{sec:discussion}.

\section{Imprecision in observables from local fluctuations}
\label{sec:error_prop}

\subsection{Propagating fluctuations from integrand to integral}
\label{sec:method1}
To demonstrate how to propagate fluctuations to an integrated quantity, suppose $K$ represents the total luminosity from a one dimensional object in a Cartesian geometry.
If this is given by $k(r)$ per unit length at position $r$, then 
\beq
K=\int_a^b k(r)\text{d}r,
\eeq
where $a$ and $b$ specify the boundary.
Now consider a local luminosity fluctuation $\delta k(r)$ with a vanishing mean but allow for inhomogeneous variance $\var{k}(r)$, i.e., $\abra{\delta k^2(r)}=\var{k}(r)$, where the angle brackets denote an ensemble average over all possible realizations of the fluctuation.
We take the fluctuations in the neighborhood of position $r$ to be coherent over a scale $l(r)$, and assume that $l$ is much smaller than the local variation scales of $k$, $\var{k}$, and $l$ itself; that is $l\ll |k/(\text{d}k/\text{d}r)|$, $l\ll |\var{k}/(\text{d}\var{k}/\text{d}r)|$, and $l\ll |l/(\text{d}l/\text{d}r)|$.
In the simplest case, for which both $\var{k}$ and $l$ are independent of position, the total luminosity becomes the sum of $N=(b-a)/l$ number of independent and identically distributed random variables, and thus the variance of $K$ is
\beq
\var{K}=\frac{\var{k}}{N}=\frac{l}{b-a}\var{k}
\eeq
according to the central limit theorem.
We now generalize to cases where both $\var{k}$ and $l$ are smooth functions of $r$. We provide two methods which give identical results. We use the first method here and a second is given in Appendix \ref{appx:method2}.

We discretize the system into grids with the size of the $i$th cell being $l_i\equiv l(r_i)$, and the grid can be constructed in the following way:
\beq
r_1=a;\ 
r_{i+1}=r_i+l_i;\ i=1,2,\cdots.
\eeq
$K$ is given by the sum of contributions from fluctuating cells,
\beq
K=\int_a^b k(r)\text{d}r\simeq\sum_i k_i l_i,
\eeq
where $k_i=k(r_i)$.
Because $l(r)\ll |k/(\text{d}k/\text{d}r)|$, 
$l_i$ can be the line element. The mean of $K^2$ is
\beq
\abra{K^2}
=\sum_{i,j}\abra{k_i k_j}l_il_j.
\eeq
Since each cell covers a local coherent length and varies independently, we have
\begin{align}
\abra{k_ik_j}
=&\left\{
\begin{aligned}
	&\abra{k_i}\abra{k_j},\ &i\neq j\\
	&\abra{k_i^2},\ &i=j
\end{aligned}
\right\}\notag\\
=&\abra{k_i}\abra{k_j}
+\delta_{ij}\left(
\abra{k_i^2}-\abra{k_i}\abra{k_j}
\right).
\end{align}
Thus
\begin{align}
\abra{K^2}
&=\sum_{i,j}\abra{k_i}\abra{k_j}l_il_j
+\sum_i \var{k_i}l_i^2\notag\\
&=\abra{K}^2+\sum_i\var{k_i}l_i^2\notag\\
&\simeq\abra{K}^2+\int_a^b \var{k}(r)l(r)\text{d}r,
\end{align}
where the last step we have used the condition $l(r)\ll |\var{k}/(\text{d}\var{k}/\text{d}r)|$.
The variance of $K$ is thus
\beq
\var{K}=\abra{K^2}-\abra{K}^2
\simeq\int_a^b\var{k}(r)l(r)\text{d}r.
 \label{eqn:varmethod1}
 \eeq

We test the analytical formula (\ref{eqn:varmethod1}) against stochastically generated data using an example 
where we have $\var{k}(r)$ and $l(r)$. Specifically 
 we take Equation (\ref{eqn:varF_DN}) for an accretion disc model, which is derived and explained in more detail in Section \ref{sec:dn}, but here this simply motivates use of the mathematical example
\beq
\var k(r)=\frac{r^{37/8}e^{0.2 r^{3/4}}}{\left(e^{0.1 r^{3/4}}-1\right)^4};\ 
l(r)=l_0r^{9/8}
\label{eqn:test16}
\eeq
with two choices $l_0=0.03$ or $l_0=0.05$.
Here length scales are in dimensionless units, normalized by the disc inner radius.
The form of $\var k(r)$ is derived based on a black-body spectrum with a mean disc surface temperature $\propto r^{-3/4}$, with $r$ being the non-dimensional disc radius.

\begin{figure}
\centering
\includegraphics[width=0.9\columnwidth]{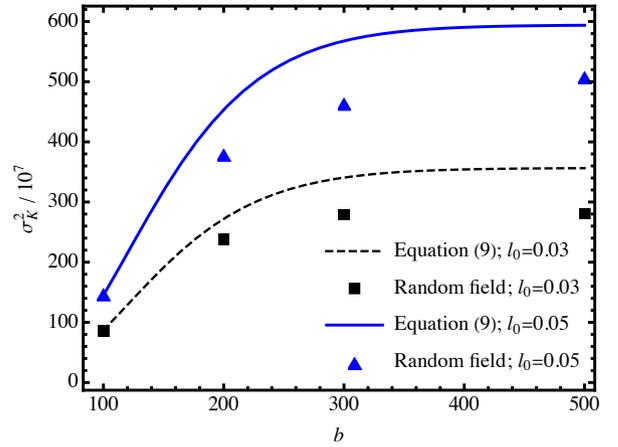}
\caption{A comparison between the variance of stochastically generated data(dots) and that from Equation (\ref{eqn:varmethod1}) (curves).}
\label{fig:var_test}
\end{figure}

Using Equation
 (\ref{eqn:test16}), 
 we compare the prediction of $\var K$ from Equation 
 (\ref{eqn:varmethod1}) with that derived from numerically generated data sets.
For the latter, we discretize the interval $[a,b]$ and assign each mesh point a random number (representing $k$) according to a multivariate Gaussian probability distribution function (PDF) with a zero mean and a covariance matrix
\beq
C_{ij}=\var{k}(r')e^{-\left[\frac{r_i-r_j}{l(r')/2}\right]^2},\ r'=\frac{r_i+r_j}{2},
\eeq
where $r_{i,j}$ are the coordinates of a pair of cells on the grid.
For each realization of $k$, its values on all mesh points are summed to give $K$, and different realizations of $k$ construct an ensemble, over which the variance of $K$ is then calculated.
With $a=1$ and different values of $b$, the results are plotted in Figure \ref{fig:var_test}.
Importantly, at a given $b$ the analytical formula is on average $\sim2000$ times faster than averaging over an ensemble with $512$ members\footnote{Both using Wolfram Mathematica on a personal computer with a quad-core CPU}, and provides an efficient way of estimating $\var K$.

Comparing the $l_0=0.03$ and $l_0=0.05$ cases,  it is evident that the larger deviation between data points and the theoretical curve at larger $b$ is due to the breakdown of the condition $l\ll|\var{K}/(\text{d}\var{K}/\text{d}r)|$.
In fact the two length scales becomes comparable at $b\gtrsim600$ with $l_0=0.03$, and $b\gtrsim400$ with $l_0=0.05$.
In general, when the correlation length becomes large enough to violate the aforementioned condition,  Equation (\ref{eqn:varmethod1}) delivers less accuate results.
In the current case and later examples, the main contribution to $\var{K}$ comes from the region where $l\ll|\var{K}/(\text{d}\var{K}/\text{d}r)|$ still holds, and therefore yields the correct order of magnitude.

Having verified Equation (\ref{eqn:varmethod1}) numerically, 
we can use it to formulate the following rules for propagating precision to an integrated quantity from its integrand:
\begin{enumerate}
\item
Replace the integrand by the variance of the integrand multiplied by the correlation length.
\item
Keep the line element of the integration unchanged.
\end{enumerate}
The result can be readily extended to included coordinate-dependent metric factors in the integration measure.

\subsection{Application to accretion discs }
Applying the formalism to an 
an axisymmetric disc, we have for the total luminosity 
\beq
K=2\pi\int_a^b k(r)r\text{d}r,
\eeq
where $k(r)$ is the luminosity per unit area from an annulus at radius $r$.
Here the line element is $\text{d}r$, and the variance of the integrand is $4\pi^2r^2\var{k}$.
Therefore
\beq
\var{K}=4\pi^2\int_a^b \var{k}(r)l(r)r^2\text{d}r.
\label{eqn:var_disc1d}
\eeq

For a two-dimensional non-axisymmetric disc, the luminosity is given by
\beq
K=\int_0^{2\pi}\text{d}\phi \int_a^b k(r,\phi)r\text{d}r.
\eeq
Assuming statistical axisymmetry, let the fluctuations of $k$ be coherent over scales of $l_r(r)$ and $l_\phi(r)$ in the radial and azimuthal directions, respectively.
The luminosity per unit length of an annulus is
\beq
K_1(r)=\int_0^{2\pi} k(r,\phi)r\text{d}\phi,
\eeq
and its variance is
\beq
\var{K_1}=\int_0^{2\pi} \var k(r)l_\phi(r)r\text{d}\phi,
\eeq
Since $K=\int K_1(r)\text{d}r$, we obtain
\beq
\var{K}=\int_0^{2\pi}\text{d}\phi \int_a^b \var{k}(r)l_r(r)l_\phi(r)r\text{d}r.
\label{eqn:var_disc2d}
\eeq
Equation (\ref{eqn:var_disc2d}) recovers the one-dimensional case (\ref{eqn:var_disc1d}) by taking $l_\phi=2\pi r$.
For quantities involving an integration in time, the method can be similarly extended.
Equation (\ref{eqn:var_disc2d}) also reflects the intuitive expectation that the variance decreases with decreasing correlation lengths because neighboring fluctuations rapidly cancel out.
For vanishing correlation lengths, 
our method is not valid and stochastic calculus must be invoked for a rigorous treatment.

Below we explore two applications of Equation (\ref{eqn:var_disc2d}), to the spectra of geometrically thin, optically thick discs.
We focus on finding spectral fluctuations at some given snapshot, where snapshot indicates a short time scale compared to eddy turnover times.

\section{Variance of predicted emission spectrum from a thermal disc}
\label{sec:spectrum}
Consider the thermal spectrum from a geometrically thin, optically thick disc.
Given some surface temperature $\Ttot(r,\phi)$ that includes both a mean and a fluctuating part, the black-body emission per unit frequency per unit area is
\beq
f(\Ttot)=\frac{2\hp \nu^3/c^2}{e^{\hp\nu/\kB \Ttot}-1},
\label{eqn:f}
\eeq
where $\hp$ is the Planck constant, $\nu$ is the photon frequency, $c$ is the speed of light, and $\kB$ is Boltzmann's constant.
The total power per unit frequency from one side of the disc is
\beq
F=\int_0^{2\pi}\text{d}\phi \int_\rin^\rout f\ r\text{d}r,
\label{eqn:F}
\eeq
where $\rin$ and $\rout$ are the inner and outer radii of the disc, respectively.

We separate the total temperature $\Ttot$ into a mean $T$ and fluctuating part $\dT$.
The mean part $T$ is assumed to be modeled by some mean-field theory which we assume to be axisymmetric and stationary so that $T=T(r)$.
The residual $\dT$ varies on turbulent time and length scales.
All fluctuations are assumed to obey statistical axisymmetry although this can be generalized.
In what follows, we give a unified formalism of calculating precision errors using the error propagation formula (\ref{eqn:var_disc2d}).
A more formal and systematic description of three important different types of errors is presented in Appendix \ref{appx:error}.

For any local fluctuation in $\Ttot$ at location $(r,\phi)$, whether from $T$ or $\dT$, with a variance $\var{\Ttot}(r)$ of its amplitude, the variance of the local fluctuation in $f$ is
\beq
\var{f}=\left[\frac{\partial f(T)}{\partial T}\right]^2\var{\Ttot}.
\eeq
As we are computing the expected variance of fluctuations around the mean, the term inside the angle brackets is evaluated at its mean value $\Ttot=T$.

We assume that the fluctuation in $f$ has the same coherent scales as those of fluctuations in $\Ttot$, as determined by the physical processes causig the latter.
If the fluctuation of $f$ is coherent over lengths $l_r(r)$ in the radial direction and $l_\phi(r)$ in the azimuthal direction, the variance of the fluctuation in $F$ will be determined by Equation (\ref{eqn:var_disc2d}) and given by
\begin{align}
\var{F}&=\int_0^{2\pi}\text{d}\phi \int_\rin^\rout
\var{f}(r) l_r(r) l_\phi(r){r\text{d}r}\notag\\
&=2\pi\int_\rin^\rout
\left(T\frac{\partial f}{\partial T}\right)^2
\frac{\var \Ttot}{T^2}
l_rl_\phi{r\text{d}r}.
\label{eqn:varF}
\end{align}
Here, $T\partial f/\partial T$ as a function of $T$ is determined by Planck's law (\ref{eqn:f}), whereas the exact expressions of $T$ itself, $\var{\Ttot}$, $l_r$, and $l_\phi$, will be model dependent.

We can also determine the fluctuation in the total luminosity for a frequency bandwidth over which fluctuations in the disc spectrum are coherent, in analogy to $l_r$ and $l_\phi$ which are coherent scales in configuration space.
Since the total luminosity is $L=2\int_0^\infty F\text{d}\nu$, the variance of its fluctuation is
\beq
\var{L}=4\int_0^\infty \var{F}\Delta \nu\text{d}\nu,
\label{eqn:varL}
\eeq
where $\Delta\nu$ is a frequency-dependent coherent bandwidth.
One such example is given in Section \ref{sec:dn}.

For convenience we define several dimensionless variables here for later use. 
The dimensionless disc radius is $\tilde r=r/\rin$.
The mean temperature can be written as $T=T_*\tilde T(\tilde r)$, where $T_*$ is the mean temperature at $\rin$ and includes all dependence on other disc parameters such as the central object mass, mass accretion rate, etc.
The dimensionless frequency is then
\beq
\beta=\frac{\hp\nu}{\kB T_*}.
\label{eqn:beta}
\eeq
We denote the disc scale height by $h(r)$, and the height-to-radius ratio at the inner boundary by $\theta=h(\rin)/\rin$.

\section{Application to protoplanetary discs}
\label{sec:pd}
Gas in protoplanetary discs is subject to turbulence, as indicated by molecular line observations \citep{Hughes2011,Guilloteau2012} and dust distributions \citep{Pinte2016}.
The underlying turbulence drivers include MRI \citep{Velikhov1959,Chandrasekhar1960,BalbusHawley1991}, self-gravitation \citep{Toomre1964,Shlosman1987}, and hydrodynamic instabilities.
The driving mechanism may also vary  between young and old objects, and between inner and outer regions, or midplane and surface layers in a single object.
Regardless of the origin, if we assume that the  eddy turnover timescale of dominant eddies is comparable to the local Keplerian timescale, the eddy turnover time can reach $\sim 30\ \text{yr}$ just at 10 AU with a central solar-mass object.
Thus for most parts of a protoplanetary disc, the turbulence timescale exceeds exposure timescales of telescopes, or even timescales of multi-epoch observations.

For such a ``snapshot'' image, some turbulent eddies are brighter and some are dimmer than the average profile predicted by a mean-field theory, and they contribute to the observed thermal spectrum at all wavelengths.
It is therefore necessary to ask, whether a  deviation between observations and theory at a specific wavelength has a truly systematic or a merely  stochastic origin.
In this section, we quantitatively incorporate the effect of turbulent fluctuations on the mean-field prediction of the disc thermal spectrum.
We isolate this turbulent effect by assuming all other parameters in the problem, e.g., mass accretion rate and the $\alpha$ parameter, remain time-independent.

Let a fluctuation in the surface temperature due to turbulence be $\dT$, which also generates a fluctuation in the luminosity $\Led$ of a turbulent eddy.
Although a   realistic probability distribution description for turbulence is still lacking,  we capture  the properties of $\dT$ 
by assuming that  the luminosity $\Led$ for each turbulent eddy,  is drawn from an ensemble with a uniform PDF,
\beq
P_{\Led}(x)=\frac{1}{2 L},\ 0,\leq x\leq 2L,
\eeq
so that $P_\Led(x)\text{d}x$ is the probability to find $\Led$ between $x$ and $x+\text{d}x$.
Here $L$ is the mean of $\Led$, and equals the luminosity of the mean-field disc model from an area equal to that of the eddy.
In general, $L$ depends on the disc radius in a mean-field model, and so does $P_\Led$.
This simple uniform PDF allows us to proceed simply, but more comprehensive statistical prescriptions may also be used \citep[e.g.,][]{LeeGammie2021}.

For optically thick discs, the PDF of $\Ttot$ can be deduced from the relation $\Led=\SB \Ttot^4$, so that
\newcommand{\Tmax}{T_\text{max}}
\beq
P_{\Ttot}(x)=\frac{2 x^3}{ \Tmax^4},\ 0\leq x\leq 2^{1/4} \Tmax,
\label{eqn:PT}
\eeq
where $\SB\Tmax^4=L$, and $\Tmax=5T/2^{9/4}$, with $T$ being the mean temperature field solved from a mean-field theory.
Equation (\ref{eqn:PT}) then defines a PDF for the disc surface temperature whose mean is $T$.
For a given mean-field disc model, $P_{\Ttot}(x)$ is known at all radii and the mean and the variance of $f$ in the ensemble can be calculated by 
\beq
\abra{f}=\int P_{\Ttot}(x)\frac{2\hp \nu^3/c^2}{e^{\hp\nu/\kB x}-1} \text{d}x
\eeq
and
\beq
\var f=\int P_{\Ttot}(x)
\left(\frac{2\hp \nu^3/c^2}{e^{\hp\nu/\kB x}-1}\right)^2\text{d}x
-\abra{f}^2,
\eeq
respectively.

Note that the ensemble mean of $f$ is different from the flux obtained by using the mean of $\Ttot$, namely
\beq
\abra{f}\neq \frac{2\hp \nu^3/c^2}{e^{\hp\nu/\kB T}-1}.
\eeq
Consequently, there is a difference between the ensemble mean of $F$, and the value that would be obtained using first the mean of $\Ttot$. The latter is what is done for  most mean-field disc models.
The difference between these two approaches leads to what we call the ``mismatch error'' (ME):
\beq
\Delta F=\int \abra{f}r\text{d}r \text{d}\phi-\left.F\right|_{\Ttot=T},
\label{eqn:ME}
\eeq
where $F$ is given by Equation (\ref{eqn:F}).

The ME reflects the disagreement between the fluxes that arise from  the following two theoretical approaches:
(i) solve for mean fields such as the mean disc temperature $T$, and then calculate the disc spectrum using $\left.F\right|_{\Ttot=T}$;
(ii) solve for both the mean and the fluctuation in temperature, the latter statistically in a mean-field model, then combine to obtain total temperature $\Ttot$ and the associated total spectrum, and  take the average to get $\abra{F}$.
Approach (i) is commonly adopted, but approach (ii) is what should be used by theorists to more accurately compare to what observations measure.

The variance of $F$ can be computed by propagating that of $f$.
The correlation length of the latter is assumed to be isotropic and identified with the turbulence scale $l$ in the model.
In an $\alpha$ disc model we have
\beq
l\simeq\frac{\alpha\cs h}{v}\simeq\frac{\alpha\Omega h^2 }{l\Omega},
\eeq
and thus $l\simeq\alpha^{1/2}h$ \citep{Blackman1998}.
The variance of $F$ is then
\beq
\var{F}=2\pi\alpha\int \var{f}h^2 r\text{d}r,
\label{eqn:varF_PD}
\eeq
using Equation (\ref{eqn:var_disc2d}).

\begin{figure}
\centering
\includegraphics[width=0.9\columnwidth]{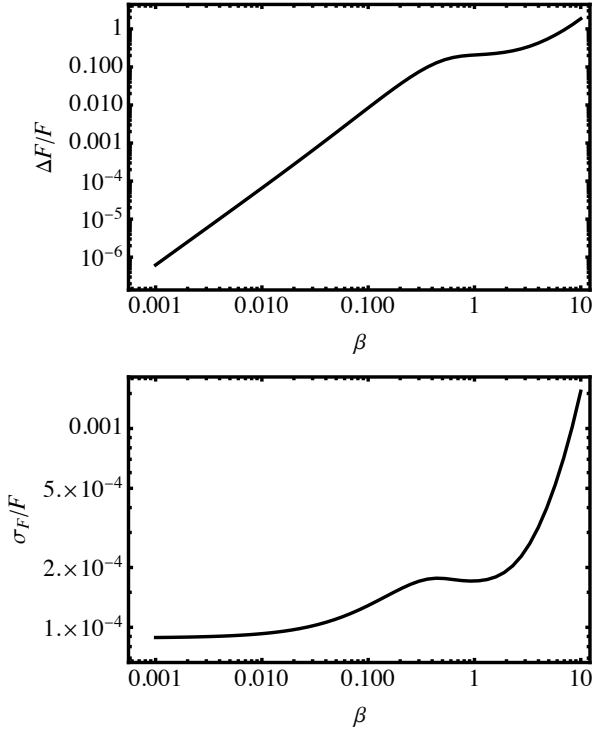}
\caption{Relative errors in the disc spectrum:
that from comparing mean of the total spectrum versus the spectrum of the mean temperature (upper),
and that from turbulent fluctuations of the surface temperature (lower).
$\beta$ is the dimensionless photon frequency defined in Equation (\ref{eqn:beta}).
The disc model uses $\alpha=0.01$, $\rout/\rin=150$, and $\theta=0.01$.
}
\label{fig:PD}
\end{figure}

We now adopt a specific profile of $T$ to qualitatively compute ME and FE.   In mean-field models the mean temperature is typically related to  the disc radius via a power-law relation, $T=T_*\tilde r^{p}$.
In the context of protoplanetary discs, $p$ varies from $-1$ to $\sim-1/2$ depending on whether disc heating is dominated by star irradiation or viscous heating.
The exponent also likely varies with the disc radius if a transition of heating source occurs.
For simplicity, we consider a constant $p$ here.
A particular model is taken from \cite{Edgar2007} for which the heat of the central plane is solely due to viscous dissipation, and the mean surface temperature $T$ and scale height $h$ are solved to be
\beq
T=T_* \tilde r^{-21/40},\ h=h_* \tilde r^{21/20}.
\label{eqn:disc_model}
\eeq
Combining Equations (\ref{eqn:ME}), (\ref{eqn:varF_PD}), and (\ref{eqn:disc_model}), we compute the ME and the FE as shown in Figure \ref{fig:PD}.

While FE originates from  turbulence  and measures the corresponding stochastic fluctuation around the mean spectrum, ME is systematic as is clear from Equation (\ref{eqn:ME}).
Consequently, fitting averaged observational data (approximately $\abra{F}$) with typical theoretical predictions that amount to $\left.F\right|_{\Ttot=T}$, introduces a bias of inferred parameters.
In the present   protoplanetary disc example
the bias in the maximum temperature at the disk inner radius $T=T(r_*)$ is $6.5\%$, that in the exponent $-\partial\ln T/\partial\ln r$ is $1.0\%$, and that for the disc outer radius is $-5.4\%$.
The numerical values are relatively small.  They do depend on the choice of the PDF for $\Ttot$, but 
the result shows that the thin disk mean field model in this context
are  quite precise, inasmuch as   observational uncertainties 
are larger.

\section{Application to dwarf novae}
\label{sec:dn}
Dwarf novae (DNe) are characterized by their regular outbursts, and 
 thought to  result from accretion disk instability \citep{Osaki1974,Hoshi1979,Lasota2001,Hameury2020}.
The standard disk instability model relies on thermal instability 
and the fact that the disc opacity changes rapidly and nonlinearly with temperature at $\sim 10^4\text{ K}$ where hydrogen ionization takes place.  
The temperature is directly connected to  the accretion rate and so  where the temperature  vs. surface density equilibrium  curve is unstable, so is the accretion rate.  Once the accretion rate increases in the disk to a value larger than the outer supply rate
can accommodate the surface density and temperature drop
until matter again builds up and the cycle repeats.  
How  exactly the disc viscosity depends  on heating is model dependent. Recently, \cite{HeldLatter2021} demonstrated that if convection results from the strong opacity 
increase, its combined effect with the magneto-rotational instability (MRI) may lead to a significant increase in angular momentum transport, characterized by cyclic bursts of $\alpha$, the stress-to-pressure parameter. The strengthened angular momentum transport  in the simulations is speculated to result from convection generated magnetic fields reseeding the MRI, an effect  most prevalent  in cases with long cooling times and short resistive times.
In these so-called strong MRI/convection cycles, $\alpha$ is enhanced in the MRI phase by approximately one order of magnitude.
Several studies have also reported similar $\alpha$ bursts in stratified shearing boxes
\citep{Simon2011,Bodo2012,Hirose2014,Coleman2018}, albeit some with different origins argued.
In this section, we build our model based on the results of \cite{HeldLatter2021}, and explore how such fluctuations in $\alpha$ can  affect the disc spectrum in the quiescent phase, during which the strong MRI/convection cycle is most likely because of the high resistivity.  

We assume that the representative temporal and spatial fluctuations in $\alpha$ are local, focusing on circumstances  that apply before  they lead to   any global coherent structure over the entire disc.
The typical cycle periods of DN bursts are observed to be $\mathcal{O}(10)$ orbital times in simulations, and therefore may occur multiple times during one hot or cold phase of the disc.
The ratio between the Keplerian timescale at the outer radius and the quiescence time of the disc can be estimated from Equations (52) or (53) in \cite{Lasota2001} as
\begin{align}
\frac{t_\text{Kep,out}}{t_\text{quiesc,oi}}\simeq&
0.0126
\left(\frac{\alpha}{0.01}\right)
\left(\frac{\dot M}{10^{17}\text{ g s}^{-1}}\right)^{2}\notag\\
&\left(\frac{\Tc}{3000\text{ K}}\right)
\left(\frac{M}{M_\odot}\right)^{0.76}
\left(\frac{\rout}{10^{10}\text{ cm}}\right)^{-4.3}
\end{align}
for an outside-in outburst, or
\begin{align}
\frac{t_\text{Kep,out}}{t_\text{quiesc,io}}\simeq&
0.00005\delta
\left(\frac{\alpha}{0.01}\right)
\left(\frac{T_c}{3000\text{ K}}\right)\notag\\
&
\left(\frac{M}{M_\odot}\right)^{-1}
\left(\frac{\rout}{10^{10}\text{ cm}}\right).
\end{align}
for an inside-out burst with large disc radii.
Here, $\Tc$ is the disc mid-plane temperature, and $\delta$ is the difference between the logarithm of the maximum surface density in the lower equilibrium branch of temperature vs. surface density during the quiescent phase;
typically $\delta\lesssim2$ \citep[c.f. Figure 11 in][]{Lasota2001}.
As such, at all radii the cycle period of the bursty $\alpha$ is much shorter than the disc quiescent time.

The relatively short cycle period leads to axisymmetric fluctuations, and accordingly the azimuthal coherence length is $l_\phi=2\pi r$.
The radial viscous diffusion time is much larger than the orbital time because
\beq
\frac{t_\text{viscous}}{t_\text{Kep}}
\simeq
\frac{r^2}{\nu_\text{T}\Omega^{-1}}
\simeq\alpha^{-1}\left(\frac{h}{r}\right)^{-2}\gg1,
\eeq
which suggests that the $\alpha$ fluctuation can also be considered local in radius.
In the simulations exhibiting the strong MRI/convection cycles \cite{HeldLatter2021}, $\alpha$ is defined by a volume average over a box of radial length $4h$, based on which we assume a coherent scale $l_r=nh$ with a fiducial $n=5$ in the radial direction.
For the standard Shakura-Sunyaev model $h/r=\theta \tilde r^{1/8}$ \citep[e.g.,][]{Frank2002book}, and thus $l_r=n\theta r \tilde r^{1/8}$.

\begin{figure}
\centering
\includegraphics[width=0.9\columnwidth]{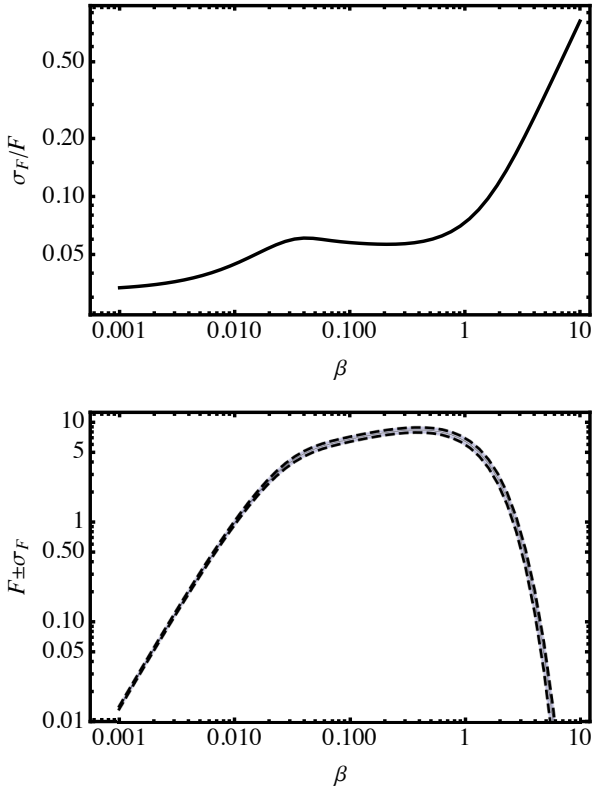}
\caption{In dimensionless units, the relative error (upper) and the corresponding error bars overplotted on the spectrum (lower) induced by $\alpha$ bursts in strong MRI/convection cycles.
$\beta$ is the dimensionless photon frequency defined in Equation (\ref{eqn:beta}).
Variation in the luminosity is $\delta L/L=0.01$.
The disc model uses $\rout/\rin=600$ and $\theta=0.01$.
}
\label{fig:DN}
\end{figure}

We now derive the temperature and flux fluctuations.
For DNe in the quiescent state, the disc is not necessarily in the global viscous equilibrium  \citep{Lasota2001}, 
but we assume that local  equilibrium between the mean surface black-body flux and the turbulent viscous dissipation holds, i.e.,
\beq
\sigma T^4(r)
=\frac{1}{2}\nu_T\Sigma\left(r\partial_r\Omega\right)^2
=\frac{1}{2}\alpha \cs h\Sigma\left(r\partial_r\Omega\right)^2.
\eeq
Since $h$, $\Sigma$, and $r\partial_r\Omega$ change on the viscous time scale, which is much longer than $10$ Keplerian orbit times, an $\alpha$ burst event that spans a timescale of $\mathcal{O}(10)$ orbits causes a mean surface temperature fluctuation  $T\propto\alpha^{1/4}$.
If the variance of fluctuations in $\alpha$ is $\var\alpha$, we have
\beq
\frac{\var{T}}{T^2}=\frac{1}{16}\frac{\var\alpha}{\alpha^2}.
\label{eqn:ie:varT}
\eeq
We  take $\sigma_\alpha/\alpha=0.5$ as estimated from \cite{HeldLatter2021}.
Since $\var{\Ttot}=\var{T}$ in this case, combining Equation (\ref{eqn:ie:varT}), $l_r=n\theta r\tilde r^{1/8}$, and $l_\phi=2\pi r$, we obtain from Equation (\ref{eqn:varF}) that
\beq
\var{F}=\frac{\pi^2 n\theta}{16}
\int_\rin^\rout\left(T\frac{\partial f}{\partial T}\right)^2
\tilde r^{1/8} r^3\text{d}r.
\label{eqn:varF_DN}
\eeq
For a standard Shakura-Sunyaev model we have $T\simeq T_*\tilde r^{-3/4}$.
The integral can be numerically carried out, and we show in Figure \ref{fig:DN} the relative fluctuation $\sigma_F/F$ and also the $1\sigma$ uncertainty around $F$ by plotting $F$ and $F\pm\sigma_{F}$ together.

To quantify the corresponding variation in the total luminosity, we need to determine  the photon frequency range over which fluctuations in the disc spectrum are  coherent.
A reasonable estimate is obtained by assuming that photons at a  frequency $\nu$ are emitted by a single annulus whose position is determined by Wien's displacement law:
\beq
\frac{\hp\nu}{\kB T(r)}\simeq 2.8
\ \Rightarrow\ 
\beta=2.8\tilr^{-3/4},
\label{eqn:Dbeta_coh}
\eeq
where $\beta=\hp\nu/k_\text{B}T_*$ is the dimensionless frequency defined.
The coherence length $l_r$ of the temperature fluctuations is propagated to a coherent frequency width $\Delta\beta_\text{coh}$ by
\beq
\Delta\beta_\text{coh}=\left|\frac{\partial \beta}{\partial \tilde r}\right|\frac{l_r}{\rin}
=2.1n\theta\tilde r^{-5/8}
=0.89n\theta\beta^{5/6}.
\eeq
One can verify that $\Delta\beta_\text{coh}<F/|\partial F/\partial\beta|$ for all $\beta\in[10^{-3},10]$.
Using Equation (\ref{eqn:varL}), the relative fluctuation in the luminosity is found to be $\delta L/L=1\%$.
If we identify the variation time $t_L$ of $L$ as that of $\alpha$ for the most luminous annulus, we find $t_L/t_\text{Kep,out}=0.0016\times\mathcal{O}(10)$ where $t_\text{Kep,out}$ is the Keplerian timescale at the disc outer radius.

\begin{figure}
\centering
\includegraphics[width=0.9\columnwidth]{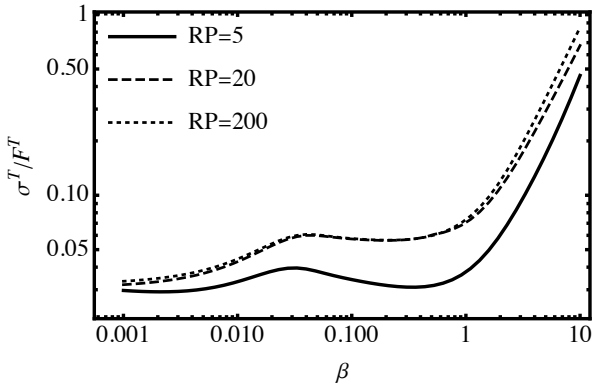}
\caption{
The same model as in Figure \ref{fig:DN}, but showing the relative error in the spectrum after it is binned by the telescope resolving power.
}
\label{fig:FT}
\end{figure}

\section{Role of telescope resolving power in determining falsifiability}
\label{sec:tel}
The imprecision we have computed above 
is independent of  the finite resolving power and binning of data by given telescope that observational data is subjected to before 
it can be compared against  theoretical predictions. 
We now show  how to incorporate this.

Let the spectral resolving power of a telescope be
\beq
\RP=\frac{\nu}{\Delta \nu},
\eeq
where $\Delta\nu$ is the telescope resolving bandwidth at photon frequency $\nu$.
For snapshot measurements, defined by  exposure time being smaller than flux fluctuation timescales, the observed flux $\FT$ is binned using the resolved bandwidth $\Delta\nu=\nu/\RP$, i.e.,
\beq
\FT(\nu)=\frac{\RP}{\nu}\int_\nu^{\nu\left(1+\RP^{-1}\right)}
F(\nu')\text{d}\nu'.
\eeq

The flux spectral fluctuations are fully  resolved when the resolved bandwidth is smaller than the coherent bandwidth of the fluctuations, i.e., when $\RP>\nu/\Delta\nu_\text{coh}$ where $\Delta\nu_\text{coh}=\Delta\beta_\text{coh}\kB T_*/\hp$.
The fluctuation in $\FT$ is then $\sigma^\text{T}=\sigma_F$ from Equation (\ref{eqn:varF_DN}).
In the opposite limit of $\RP< \nu/\Delta\nu_\text{coh}$, every neighboring number $\Delta\nu/\Delta\nu_\text{coh}$  of fluctuating bandwidths are binned.
Identifying $K$ with $\FT$ and $k$ with $F$ in Equation (\ref{eqn:var_disc2d}) 
we have
\beq
\sigma^\text{T}
\simeq \frac{\RP}{\nu}
\left[\int_\nu^{\nu\left(1+\RP^{-1}\right)}
\sigma^2(\nu')\Delta \nu'_\text{coh}\text{d}\nu'
\right]^{1/2}.
\label{eqn:varFT}
\eeq

In Figure \ref{fig:FT} we plot the relative precision $\sigma^T/\FT$ for different constant $\RP$ values by evaluating Equation (\ref{eqn:varFT}).
The critical dimensionless photon frequencies above which $\RP>\beta/\Delta\beta_\text{coh}$ are $10^{-3.9}$, $10^{-0.3}$, and $10^{5.7}$ for the $\RP=5,20,200$ cases, respectively.
Smaller RP reduces the effect of fluctuations because more fluctuation cells are averaged within a single resolving bandwidth.

\section{Conclusion}
\label{sec:discussion}
The scale separations between mean and fluctuating fields in astrophysical flows are finite, in contrast to the idealized assumption of infinity, which affects both the accuracy and the precision of a given mean-field theory.
In particular, spatially or temporally averaged observational data unavoidably includes contributions from small-scale fluctuations, and thus when fitting data, the inferred model may fail to match those from an accurate theory, or misleadingly appear to agree with an inaccurate theory, if fluctuations around mean-field results are not properly estimated.
Falsifiability of mean-field theories by comparing to observations requires careful distinction between disagreements that result from accuracy with those that result from imprecision.

While improving accuracy means increasing the fidelity of the input physics that account for the finite scale separation, in this work we focus on calculations of precision of mean-field theories, defined as the variance of the fluctuations of mean-fields propagated from small-scale fields.
We have considered the general case of statistically inhomogeneous small-scale fluctuations, and the derived Equation (\ref{eqn:varmethod1}) is shown to be consistent with data from a numerically realized ensemble, yet significantly preceding the latter in terms of efficiency.

We then exemplified the method by computing imprecision in the prediction of accretion disc thermal spectra, induced by (i) turbulent fluctuations, and (ii) meso-scale fluctuations in $\alpha$, respectively.
For both, we imposed fluctuations in temperature and its coherence length allowing both to be local functions of space.
The consequent error propagation of these fluctuations to global observables, namely spectra and luminosity, were then derived.
Although with small magnitudes, the derived spectra fluctuations indeed depend on photon frequency, and suggests that accurate falsification of a mean-field theory may require a likelihood function or data weighting that reflects the fidelity of the theoretical mean-field values in different regions of the parameter space, as the photon frequency in the current examples.

In contrast to previous dynamical models \citep{Lyubarskii1997,DexterAgol2011,CowperthwiteReynolds2014,TurnerReynolds2021}, each of which considers  different models for the time evolution of imposed surface density or temperature fluctuations and pursues their observational signature, our work focuses on an efficient semi-analytical method of computing the propagation of inhomogeneous fluctuations to synthetic observables to compare with snapshot observations.
We have assessed how precise the predictions of standard mean-field disc theories are when subjected to such fluctuations of a given amplitude.

In addition to the stochastic contributions, we have also shown that the nonlinear relation between basic physical quantities solved in theories (i.e., surface temperature) and observables (i.e., disc spectrum) can lead to a finite systematic mismatch between the meaning of the quantity that is predicted and that is observed and then averaged.
This leads to a systematic bias when backing out disc parameters of a few per cent, and a more comprehensive treatment of turbulence would offer a more accurate determination of this bias.

\section*{Acknowledgements}
We thank Jing Yang and Alexandra Veledina for useful discussions, and we also thank the referees for helpful suggestions.
HZ acknowledges support from Horton Fellowship from the Laboratory for Laser Energetics at U. Rochester.
EB acknowledges support from NSF grants AST-1813298, PHY-2020249; DOE grants DE-SC0020432 and DE-SC0020103; KITP UC Santa Barbara funded by NSF Grant PHY-1748958.

\appendix
\section{Second Method to derive Equation (9)}
\label{appx:method2}
We construct intervals on $[a,b]$ with scale $c(r)$ such that $l(r)\ll c(r)$ and in each interval, $\var{k}$ and $l$ are approximately constant.
Consider a coarse-grained field
\beq
k'(r)=\int_a^b k(r-s)G_{c(r)}(s)\text{d}s
\eeq
and the corresponding integral
\beq
K'=\int_a^b k'(r)\text{d}r.
\eeq
Here $G_c(s)$ is a normalized kernel with compact support from $|s|\leq c$ where $c$ is a function of $r$, satisfying $l(r)\ll c(r)\ll |k/(\text{d} k /\text{d}r)|$ and $l(r)\ll c(r)\ll |\var{k}/(\text{d}\var{k}/\text{d}r)|$.
We then have
\begin{align}
K'
=&\int_a^b G_{c(r)}(s)\text{d}s
\int_a^b k(r-s)\text{d}r\notag\\
=&\int_a^b G_{c(r)}(s)\text{d}s
\int_{a-s}^{b-s} k(r)\text{d}r\notag\\
\simeq&
\int_a^b G_{c(r)}(s)\text{d}s
\int_a^b k(r)\text{d}r
=K.
\end{align}
The last equality follows the normalization property of $G$.
Thus, we only need to quantify the variance of $K'$.

The coarse-grained field $k'$ locally has a Gaussian distribution because of the central limit theorem.
For each averaging cell of size $c(r)$, the number of fluctuating cells of size $l(r)$ is $c(r)/l(r)$, and therefore
\beq
\var{k'}=\frac{\var{k}}{c(r)/l(r)}.
\eeq

Now we divide the entire region region $a\leq r\leq b$ into grids, constructed by
\beq
r_1=a,\ 
r_{i+1}=r_i+c_i,\ i=1,2,\cdots,
\eeq
where $c_i=c(r_i)$.
Since $c(r)\ll |k/(\text{d} k /\text{d}r)|$, $K'$ can be approximated by
\beq
K'=\sum_i k'_i c_i,
\eeq
and since each $k'_i$ is a Gaussian, the variance of $K'$, which is equal to that of $K$, is given by
\begin{align}
&\var{K}=
\var{K'}\notag\\
=&\sum_i \var{k'_i}c_i^2
=\sum_i \frac{\var{k_i}}{c_i/l_i} c_i^2
=\sum_i \var{k_i}l_i c_i\notag\\
\simeq& \int_a^b \var{k}(r)l(r)\text{d}r,
\label{eqn:var_1d}
\end{align}
which agrees with the derivation in Section \ref{sec:method1}.

\section{A formal discussion of different types of errors}
\label{appx:error}
In this Appendix we present a formal discussion of possible deviations of mean-field theory predictions from observed values, assuming that observations could be made with infinite resolution.

To clarify the idea, we define a snapshot measurement $\Fobs$ of some observable from an object that possesses turbulent fluctuations.
We also consider a multi-epoch average (with inter-epoch time longer than the turbulent timescale) of snapshots, assuming each to be drawn from an ensemble $\Dobs$ of identical discs with different turbulence realizations and turbulence being ergodic.
Then we consider this multi-epoch average to be an ensemble average $\abraobs\Fobs$;
the superscript of the angle brackets means the average is drawn from the ensemble $\Dobs$.

Now suppose that we have a theoretical model that predicts mean-field quantities from the same, yet {\it theoretical}, turbulent object.
We call the ensemble associated with this theoretical turbulent object $\Dth$.
How well $\Dobs$ coincides with $\Dth$ defines the {\it accuracy} of the theory.
Improving accuracy means increasing the fidelity of the input physics.
Here we assume that the theoretical mean-field model is fully accurate and instead focus on its {\it precision}.

For a member of $\Dth$, the theory predicts a mean value of the observable, $\Fth$.
For a mean-field theory, $\Fth$ is the same for all members of $\Dth$, equal to its ensemble average in $\Dth$: $\abrath\Fth=\Fth$;
here, the superscript of the angle brackets means the average is drawn from $\Dth$.
From each element of $\Dth$, we may construct synthetic observations 
which represent specific predictions processed from the mean-field theory to match what a 
given telescope would measure. We denote this by $\Fobsth$.
Note that $\Fobsth$ represents a predicted snapshot measurement of a turbulent object and so we can also construct 
 the ensemble average $\abrath\Fobsth$.
The two quantities $\Fth$ and $\abrath\Fobsth$ differ 
in general, and indeed differed for the example model in Section \ref{sec:pd}.

Having clarified the notation, the difference between an actual observed value
and theoretical mean-field value for the observable is
\begin{align}
\dF= & \Fobs-\Fth\notag\\
\equiv & \Fobs-\abraobs\Fobs
\label{eqn:dF_FE}\\
&+\abraobs\Fobs-\abrath\Fobsth
\label{eqn:dF_accuracy}\\
&+\abrath\Fobsth-\Fth
\label{eqn:dF_ME}, 
\end{align}
where we have decomposed the right side into three differences:
Difference (\ref{eqn:dF_FE}) is how much a real observation deviates from its multi-epoch average;
difference (\ref{eqn:dF_accuracy}) vanishes if the two ensembles $\Dth$ and $\Dobs$ are identical, quantifying the accuracy of the theory;
difference (\ref{eqn:dF_ME}) measures whether the quantity predicted ($\Fth$) means the same thing as the quantity observations actually measure ($\abrath\Fobsth$).
We have assumed an accurate theory, so we set 
\beq
\Fobs=\Fobsth.
\eeq
Furthermore, we need not distinguish in which ensemble fields are averaged, and use $\abraobs\cdot=\abrath\cdot=\abra\cdot$ in what follows.
Consequently, Term (\ref{eqn:dF_accuracy}) vanishes.

Since $\dF$ fluctuates in the ensemble, it will be meaningful to quantify its mean $\abra\dF$ and variance $\var\dF$.
Using Terms (\ref{eqn:dF_FE}) to (\ref{eqn:dF_ME}),
we define the ``filtering error'' (FE) as
\beq
\FE^2=\var{\Fobs-\abra\Fobs}=\var\Fobs=\var\Fobsth
\label{eqn:def_FE}
\eeq
where $\var X$ denotes the variance of the quantity $X$ in the ensemble, and the last step comes from our accuracy assumption $\Fobs=\Fobsth$.
The ``mismatch error'' (ME) is
\beq
\Delta F=\abra{\abra\Fobsth-\Fth}=\abra\Fobsth-\Fth.
\label{eqn:def_ME}
\eeq
In addition, the error associated with the perturbations in $\Fth$ induced by varying model parameters (e.g., boundary conditions, transport coefficients) contributes an ``intrinsic error'' (IE), $\IE^2$.
We then have
\begin{align}
&\abra\dF=\Delta F,\\
&\var\dF=\IE^2+\FE^2.
\end{align}

The IE, FE and ME can all be determined theoretically because they only involve $\Fth$ and $\Fobsth$, but not $\Fobs$.
Correspondingly, $\Fth+\abra\dF\pm\sigma_{\dF}$ will be a mean-field prediction with error bars, giving a finite range of where we expect the observed value $\Fobs$ to locate.
Thus to facilitate a more appropriate comparison between theory and observations than what is commonly 
done, we must quantify the precision of the former so that the error can be added to the mean and the result compared against an observation when that obsevation is a member of an ensemble rather than an ensemble average.

The non-vanishing ME in thermal disc spectra can be elucidated in the following way.
Consider a member from $\Dth$.
Let its surface temperature, as if we could measure it, be $\Tobsth$, which is turbulent and varies with the disc radius $r$ and azimuthal position $\phi$.
The observed total power emitted per unit frequency from one side of the disc is
\begin{align}
\Fobsth&=\int_0^{2\pi}\text{d}\phi 
\int_\rin^\rout
\frac{2\hp\nu^3/c^2}{e^{\hp\nu/\kB\Tobsth}-1}
{r\text{d}r}\notag\\
&\equiv
I\left[\Tobsth\right],
\end{align}
where $I$ is the spectrum functional.
The ensemble mean of the observation is $\abra{\Fobsth}$.
On the other hand, in a mean-field disc theory, the constructed equations are solved to obtain a mean surface temperature $\bTs$.
The predicted total emission is
\beq
\Fth=I\left[\bTs\right]
\eeq
Even assuming an accurate theory $\bTs=\abra\Tobsth$, we still have
\beq
\abra{\Fobsth}=\abra{I\left[\Tobsth\right]}
\neq I\left[\abra\Tobsth\right]
=\Fth
\eeq
because of the nonlinear dependence on the surface temperature.
The difference between the left and right sides is defined as the ME.

\section*{Data availability}
The data underlying this article will be shared on reasonable request to the corresponding author.


\bibliographystyle{mnras}
\bibliography{refs}

\end{document}

%% file: macro.tex
\newcommand{\beq}{\begin{equation}}
\newcommand{\eeq}{\end{equation}}

\newcommand{\benum}{\begin{enumerate}}
\newcommand{\eenum}{\end{enumerate}}

\def\apj{ApJ}

\def\aap{A \& Ap}

\def\pasj{PASJ}

\def\nat{Nature}
\def\mnras{MNRAS}
\def\apj{ApJ}
\def\apjl{ApJL}
\def\aap{A\&A}

\newcommand{\abra}[1]{\left\langle{#1}\right\rangle}

\newcommand{\cs}{{c_\text{S}}}

\newcommand{\rin}{{r_*}}
\newcommand{\rout}{{r_\text{out}}}

\renewcommand{\var}[1]{\sigma^2_{#1}}

\newcommand{\Led}{{L_\text{ed}}}
\newcommand{\IE}{\sigma_\text{IE}}
\newcommand{\FE}{\sigma_\text{FE}}
\newcommand{\bTs}{T^\text{th}}
\newcommand{\kB}{k_\text{B}}

\newcommand{\Fobs}{{F^\text{obs}}}
\newcommand{\Fth}{{F^\text{th}}}
\newcommand{\dF}{{\delta F}}

\newcommand{\hp}{{h_\text{P}}}

\newcommand{\tilr}{{\tilde r}}
\newcommand{\Tc}{{T_\text{c}}}

\newcommand{\SB}{{\sigma_\text{SB}}}
\newcommand{\RP}{\text{RP}}

\newcommand{\Dobs}{\mathcal{D}^\text{obs}}
\newcommand{\Dth}{\mathcal{D}^\text{th}}
\newcommand{\abrath}[1]{\left\langle{#1}\right\rangle^\text{th}}
\newcommand{\abraobs}[1]{\left\langle{#1}\right\rangle^\text{obs}}
\newcommand{\Fobsth}{F^\text{obs,th}}

\newcommand{\Tobsth}{T^\text{obs,th}}
\newcommand{\FT}{F^\text{T}}

\newcommand{\Ttot}{\mathscr{T}}

\newcommand{\dT}{\delta\Ttot}

%% file: discPrecision.bbl
\begin{thebibliography}{}
\makeatletter
\relax
\def\mn@urlcharsother{\let\do\@makeother \do\$\do\&\do\#\do\^\do\_\do\%\do\~}
\def\mn@doi{\begingroup\mn@urlcharsother \@ifnextchar [ {\mn@doi@}
  {\mn@doi@[]}}
\def\mn@doi@[#1]#2{\def\@tempa{#1}\ifx\@tempa\@empty \href
  {http://dx.doi.org/#2} {doi:#2}\else \href {http://dx.doi.org/#2} {#1}\fi
  \endgroup}
\def\mn@eprint#1#2{\mn@eprint@#1:#2::\@nil}
\def\mn@eprint@arXiv#1{\href {http://arxiv.org/abs/#1} {{\tt arXiv:#1}}}
\def\mn@eprint@dblp#1{\href {http://dblp.uni-trier.de/rec/bibtex/#1.xml}
  {dblp:#1}}
\def\mn@eprint@#1:#2:#3:#4\@nil{\def\@tempa {#1}\def\@tempb {#2}\def\@tempc
  {#3}\ifx \@tempc \@empty \let \@tempc \@tempb \let \@tempb \@tempa \fi \ifx
  \@tempb \@empty \def\@tempb {arXiv}\fi \@ifundefined
  {mn@eprint@\@tempb}{\@tempb:\@tempc}{\expandafter \expandafter \csname
  mn@eprint@\@tempb\endcsname \expandafter{\@tempc}}}

\bibitem[\protect\citeauthoryear{{Balbus} \& {Hawley}}{{Balbus} \&
  {Hawley}}{1991}]{BalbusHawley1991}
{Balbus} S.~A.,  {Hawley} J.~F.,  1991, \mn@doi [\apj] {10.1086/170270}, \href
  {https://ui.adsabs.harvard.edu/abs/1991ApJ...376..214B} {376, 214}

\bibitem[\protect\citeauthoryear{{Blackman}}{{Blackman}}{1998}]{Blackman1998}
{Blackman} E.~G.,  1998, \mn@doi [\mnras] {10.1046/j.1365-8711.1998.01967.x},
  \href {https://ui.adsabs.harvard.edu/abs/1998MNRAS.299L..48B} {299, L48}

\bibitem[\protect\citeauthoryear{{Blackman}, {Nauman}  \& {Edgar}}{{Blackman}
  et~al.}{2010}]{Blackman2010}
{Blackman} E.~G.,  {Nauman} F.,   {Edgar} R.~G.,  2010, arXiv e-prints, \href
  {https://ui.adsabs.harvard.edu/abs/2010arXiv1010.1478B} {p. arXiv:1010.1478}

\bibitem[\protect\citeauthoryear{{Bodo}, {Cattaneo}, {Mignone}  \&
  {Rossi}}{{Bodo} et~al.}{2012}]{Bodo2012}
{Bodo} G.,  {Cattaneo} F.,  {Mignone} A.,   {Rossi} P.,  2012, \mn@doi [\apj]
  {10.1088/0004-637X/761/2/116}, \href
  {https://ui.adsabs.harvard.edu/abs/2012ApJ...761..116B} {761, 116}

\bibitem[\protect\citeauthoryear{{Chandrasekhar}}{{Chandrasekhar}}{1960}]{Chandrasekhar1960}
{Chandrasekhar} S.,  1960, \mn@doi [Proceedings of the National Academy of
  Science] {10.1073/pnas.46.2.253}, \href
  {https://ui.adsabs.harvard.edu/abs/1960PNAS...46..253C} {46, 253}

\bibitem[\protect\citeauthoryear{{Coleman}, {Blaes}, {Hirose}  \&
  {Hauschildt}}{{Coleman} et~al.}{2018}]{Coleman2018}
{Coleman} M. S.~B.,  {Blaes} O.,  {Hirose} S.,   {Hauschildt} P.~H.,  2018,
  \mn@doi [\apj] {10.3847/1538-4357/aab6a7}, \href
  {https://ui.adsabs.harvard.edu/abs/2018ApJ...857...52C} {857, 52}

\bibitem[\protect\citeauthoryear{{Cowperthwaite} \& {Reynolds}}{{Cowperthwaite}
  \& {Reynolds}}{2014}]{CowperthwiteReynolds2014}
{Cowperthwaite} P.~S.,  {Reynolds} C.~S.,  2014, \mn@doi [\apj]
  {10.1088/0004-637X/791/2/126}, \href
  {https://ui.adsabs.harvard.edu/abs/2014ApJ...791..126C} {791, 126}

\bibitem[\protect\citeauthoryear{{Cox} \& {Giuli}}{{Cox} \&
  {Giuli}}{1968}]{CoxGiuli1968}
{Cox} J.~P.,  {Giuli} R.~T.,  1968, {Principles of stellar structure}

\bibitem[\protect\citeauthoryear{{Dexter} \& {Agol}}{{Dexter} \&
  {Agol}}{2011}]{DexterAgol2011}
{Dexter} J.,  {Agol} E.,  2011, \mn@doi [\apjl] {10.1088/2041-8205/727/1/L24},
  \href {https://ui.adsabs.harvard.edu/abs/2011ApJ...727L..24D} {727, L24}

\bibitem[\protect\citeauthoryear{{Edgar}, {Quillen}  \& {Park}}{{Edgar}
  et~al.}{2007}]{Edgar2007}
{Edgar} R.~G.,  {Quillen} A.~C.,   {Park} J.,  2007, \mn@doi [\mnras]
  {10.1111/j.1365-2966.2007.12305.x}, \href
  {https://ui.adsabs.harvard.edu/abs/2007MNRAS.381.1280E} {381, 1280}

\bibitem[\protect\citeauthoryear{{Frank}, {King}  \& {Raine}}{{Frank}
  et~al.}{2002}]{Frank2002book}
{Frank} J.,  {King} A.,   {Raine} D.~J.,  2002, {Accretion Power in
  Astrophysics: Third Edition}

\bibitem[\protect\citeauthoryear{{Guilloteau}, {Dutrey}, {Wakelam}, {Hersant},
  {Semenov}, {Chapillon}, {Henning}  \& {Pi{\'e}tu}}{{Guilloteau}
  et~al.}{2012}]{Guilloteau2012}
{Guilloteau} S.,  {Dutrey} A.,  {Wakelam} V.,  {Hersant} F.,  {Semenov} D.,
  {Chapillon} E.,  {Henning} T.,   {Pi{\'e}tu} V.,  2012, \mn@doi [\aap]
  {10.1051/0004-6361/201220331}, \href
  {https://ui.adsabs.harvard.edu/abs/2012A&A...548A..70G} {548, A70}

\bibitem[\protect\citeauthoryear{{Hameury}}{{Hameury}}{2020}]{Hameury2020}
{Hameury} J.~M.,  2020, \mn@doi [Advances in Space Research]
  {10.1016/j.asr.2019.10.022}, \href
  {https://ui.adsabs.harvard.edu/abs/2020AdSpR..66.1004H} {66, 1004}

\bibitem[\protect\citeauthoryear{{Held} \& {Latter}}{{Held} \&
  {Latter}}{2021}]{HeldLatter2021}
{Held} L.~E.,  {Latter} H.~N.,  2021, \mn@doi [\mnras] {10.1093/mnras/stab974},
  \href {https://ui.adsabs.harvard.edu/abs/2021MNRAS.tmp..985H} {}

\bibitem[\protect\citeauthoryear{{Hirose}, {Blaes}, {Krolik}, {Coleman}  \&
  {Sano}}{{Hirose} et~al.}{2014}]{Hirose2014}
{Hirose} S.,  {Blaes} O.,  {Krolik} J.~H.,  {Coleman} M. S.~B.,   {Sano} T.,
  2014, \mn@doi [\apj] {10.1088/0004-637X/787/1/1}, \href
  {https://ui.adsabs.harvard.edu/abs/2014ApJ...787....1H} {787, 1}

\bibitem[\protect\citeauthoryear{{H{\={o}}shi}}{{H{\={o}}shi}}{1979}]{Hoshi1979}
{H{\={o}}shi} R.,  1979, \mn@doi [Progress of Theoretical Physics]
  {10.1143/PTP.61.1307}, \href
  {https://ui.adsabs.harvard.edu/abs/1979PThPh..61.1307H} {61, 1307}

\bibitem[\protect\citeauthoryear{{Hughes}, {Wilner}, {Andrews}, {Qi}  \&
  {Hogerheijde}}{{Hughes} et~al.}{2011}]{Hughes2011}
{Hughes} A.~M.,  {Wilner} D.~J.,  {Andrews} S.~M.,  {Qi} C.,   {Hogerheijde}
  M.~R.,  2011, \mn@doi [\apj] {10.1088/0004-637X/727/2/85}, \href
  {https://ui.adsabs.harvard.edu/abs/2011ApJ...727...85H} {727, 85}

\bibitem[\protect\citeauthoryear{{Krause} \& {Raedler}}{{Krause} \&
  {Raedler}}{1980}]{KrauseRaedler1980}
{Krause} F.,  {Raedler} K.-H.,  1980, {Mean-field magnetohydrodynamics and
  dynamo theory}

\bibitem[\protect\citeauthoryear{{Lasota}}{{Lasota}}{2001}]{Lasota2001}
{Lasota} J.-P.,  2001, \mn@doi [\nar] {10.1016/S1387-6473(01)00112-9}, \href
  {https://ui.adsabs.harvard.edu/abs/2001NewAR..45..449L} {45, 449}

\bibitem[\protect\citeauthoryear{{Lee} \& {Gammie}}{{Lee} \&
  {Gammie}}{2021}]{LeeGammie2021}
{Lee} D.,  {Gammie} C.~F.,  2021, \mn@doi [\apj] {10.3847/1538-4357/abc8f3},
  \href {https://ui.adsabs.harvard.edu/abs/2021ApJ...906...39L} {906, 39}

\bibitem[\protect\citeauthoryear{{Lyubarskii}}{{Lyubarskii}}{1997}]{Lyubarskii1997}
{Lyubarskii} Y.~E.,  1997, \mn@doi [\mnras] {10.1093/mnras/292.3.679}, \href
  {https://ui.adsabs.harvard.edu/abs/1997MNRAS.292..679L} {292, 679}

\bibitem[\protect\citeauthoryear{{Osaki}}{{Osaki}}{1974}]{Osaki1974}
{Osaki} Y.,  1974, \pasj, \href
  {https://ui.adsabs.harvard.edu/abs/1974PASJ...26..429O} {26, 429}

\bibitem[\protect\citeauthoryear{{Pinte}, {Dent}, {M{\'e}nard}, {Hales},
  {Hill}, {Cortes}  \& {de Gregorio-Monsalvo}}{{Pinte}
  et~al.}{2016}]{Pinte2016}
{Pinte} C.,  {Dent} W.~R.~F.,  {M{\'e}nard} F.,  {Hales} A.,  {Hill} T.,
  {Cortes} P.,   {de Gregorio-Monsalvo} I.,  2016, \mn@doi [\apj]
  {10.3847/0004-637X/816/1/25}, \href
  {https://ui.adsabs.harvard.edu/abs/2016ApJ...816...25P} {816, 25}

\bibitem[\protect\citeauthoryear{{Roberts} \& {Soward}}{{Roberts} \&
  {Soward}}{1975}]{RobertsSoward1975}
{Roberts} P.~H.,  {Soward} A.~M.,  1975, \mn@doi [Astronomische Nachrichten]
  {10.1002/asna.19752960202}, \href
  {http://adsabs.harvard.edu/abs/1975AN....296...49R} {296, 49}

\bibitem[\protect\citeauthoryear{{Shakura} \& {Sunyaev}}{{Shakura} \&
  {Sunyaev}}{1973}]{SS73}
{Shakura} N.~I.,  {Sunyaev} R.~A.,  1973, \aap, \href
  {https://ui.adsabs.harvard.edu/abs/1973A&A....24..337S} {500, 33}

\bibitem[\protect\citeauthoryear{{Shlosman} \& {Begelman}}{{Shlosman} \&
  {Begelman}}{1987}]{Shlosman1987}
{Shlosman} I.,  {Begelman} M.~C.,  1987, \mn@doi [\nat] {10.1038/329810a0},
  \href {https://ui.adsabs.harvard.edu/abs/1987Natur.329..810S} {329, 810}

\bibitem[\protect\citeauthoryear{{Simon}, {Hawley}  \& {Beckwith}}{{Simon}
  et~al.}{2011}]{Simon2011}
{Simon} J.~B.,  {Hawley} J.~F.,   {Beckwith} K.,  2011, \mn@doi [\apj]
  {10.1088/0004-637X/730/2/94}, \href
  {https://ui.adsabs.harvard.edu/abs/2011ApJ...730...94S} {730, 94}

\bibitem[\protect\citeauthoryear{{Toomre}}{{Toomre}}{1964}]{Toomre1964}
{Toomre} A.,  1964, \mn@doi [\apj] {10.1086/147861}, \href
  {http://adsabs.harvard.edu/abs/1964ApJ...139.1217T} {139, 1217}

\bibitem[\protect\citeauthoryear{{Turner} \& {Reynolds}}{{Turner} \&
  {Reynolds}}{2021}]{TurnerReynolds2021}
{Turner} S. G.~D.,  {Reynolds} C.~S.,  2021, \mn@doi [\mnras]
  {10.1093/mnras/stab875}, \href
  {https://ui.adsabs.harvard.edu/abs/2021MNRAS.504..469T} {504, 469}

\bibitem[\protect\citeauthoryear{{Velikhov}}{{Velikhov}}{1959}]{Velikhov1959}
{Velikhov} E.~P.,  1959, Sov. Phys. JETP, 36, 995

\bibitem[\protect\citeauthoryear{{Zhou}, {Blackman}  \& {Chamandy}}{{Zhou}
  et~al.}{2018}]{ZhouBlackmanChamandy2018}
{Zhou} H.,  {Blackman} E.~G.,   {Chamandy} L.,  2018, \mn@doi [Journal of
  Plasma Physics] {10.1017/S0022377818000375}, \href
  {http://adsabs.harvard.edu/abs/2018JPlPh..84c7302Z} {84, 735840302}

\makeatother
\end{thebibliography}
